\documentclass[aps,twocolumn,showpacs]{revtex4}
\usepackage{graphicx}
\begin{document}
\bibliographystyle{revtex}
\title{Toroidal, compression, and vortical dipole strengths in $^{144-154}$Sm:
\\
Skyrme-RPA exploration of deformation effect}
\author{J. Kvasil$^1$, V.O. Nesterenko$^2$,  W. Kleinig$^{2,3}$, D. Bozik$^1$,
P.-G. Reinhard$^4$, and N. Lo Iudice$^{5,6}$,} \affiliation{$^1$
Institute of Particle and Nuclear Physics, Charles University,
CZ-18000, Praha 8, Czech Republic}
\email{kvasil@ipnp.troja.mff.cuni.cz} \affiliation{$^2$ Laboratory
of Theoretical Physics, Joint Institute for Nuclear Research,
Dubna, Moscow region, 141980, Russia} \email{nester@theor.jinr.ru}
\affiliation{$^3$ Technische Universit\"at Dresden, Inst. f\"ur
Analysis,D-01062, Dresden, Germany} \affiliation{$^4$ Institut
f\"ur Theoretische Physik II, Universit\"at Erlangen, D-91058,
Erlangen, Germany} \affiliation{$^5$ Dipartimento di Scienze
Fisiche, Universit$\grave{\rm{a}}$ di Napoli  Federico II, Monte S
Angelo, Via Cintia I-80126 Napoli, Italy} \affiliation{$^6$
Istituto Nazionale di Fisica Nucleare, Sezione di Napoli, Monte S
Angelo, Monte S Angelo, Via Cintia I-80126 Napoli, Italy}

\date{\today}

\begin{abstract}
A comparative analysis of toroidal, compressional and vortical dipole
strengths in the spherical $^{144}$Sm and the deformed $^{154}$Sm is
performed within the random-phase-approximation using a set of
different Skyrme forces.  Isoscalar (T=0), isovector (T=1), and
electromagnetic excitation channels are considered.  The role of the
nuclear convection $j_{\text{con}}$ and magnetization $j_{\text{mag}}$
currents is inspected. It is shown that the deformation leads to an
appreciable redistribution of the strengths and causes
a spectacular deformation splitting (exceeding 5 MeV) of the isoscalar compressional
mode.  In $^{154}$Sm, the $\mu$=0 and $\mu$=1
branches of the mode form well separated resonances. When stepping
from $^{144}$Sm to $^{154}$Sm, we observe an increase of the toroidal,
compression and vortical contributions in the low-energy region (often
called pygmy resonance). The strength in this region seems to be an
overlap of various excitation modes. The energy centroids of the
strengths depend significantly on the isoscalar effective mass $m_0$.
Skyrme forces with a large $m_0$ (typically $m_0/m \approx 0.8 - 1$)
seem to be more suitable for description of experimental data for the
isoscalar giant dipole resonance.
\end{abstract}

\pacs{24.30.Cz,21.60.Jz,13.40.-f,27.80.+w}

\maketitle

\section{Introduction}

Possible evidence of a vortical flow in nuclei was suggested by
the analysis of the structure of the isoscalar giant dipole resonance
(ISGDR) observed in $(\alpha, \alpha')$ scattering experiments
\cite{Clark01,Youngb04,Itoh_PRC_03,Uchi03,Uchi04} (see
Ref. \cite{Pa07} for a review and an exhaustive list of references).
In fact, the prevailing conclusion from this analysis was that the
high-energy peak of the ISGDR is produced by compressional vibrations
\cite{Vret00,Colo00}, whereas the low-energy bump should be attributed
to vortical nuclear flow associated with a toroidal dipole mode
\cite{Bast93,Vret02,Kv03,Mis06}.  The toroidal moments emerge as
second order terms of multipole expansions of electric currents in
physical systems \cite{Dub75,Dub83}. In nuclei, the toroidal mode (TM)
was predicted within a hydrodynamical model \cite{Sem81}.

On the other hand, it was argued that a strong mixing between
compressional and vortical vibrations in the isoscalar E1 states
should be expected \cite{Mis06}. Moreover, it is not yet completely
settled how the vorticity relates to the toroidal and compressional
modes. This depends on the way the vorticity is defined. In hydrodynamical
(HD) models, where it is characterized by a non-vanishing curl of the
velocity field \cite{La87}, the vorticity is solely associated to the
TM while the compressional mode (CM) is irrotational. An alternative
definition, more linked to nuclear observables, was proposed by Ravenhall
and Wambach (RW) \cite{Ra87}. It adopts, as a measure of the
vorticity, the multipole component $j^{(fi)}_{\lambda l=\lambda+1}(r)$
of the transition density current $\langle
f|\hat{\vec{j}}_{\rm{nuc}}(\vec{r}) |i\rangle$. The motivation for this
choice is that this component is not constrained by the continuity
equation. With this measure of the vorticity, the TM and CM describe mixed
flows of both vortical and irrotational nature.
In the recent study \cite{Kv11}, the vortical operator of RW type
was derived and related in a simple manner to the CM and TM
operators. Then the vortical, toroidal and compression E1 strengths
in $^{208}$Pb were compared and thoroughly scrutinized  within a
separable random-phase-approximation (SRPA) \cite{Ne02,Ne06} using
the SLy6 Skyrme force \cite{Ch97}. Later a similar study was
performed for the isotopes $^{100,124,132}$Sn \cite{Sn_PS_13}.

Following the above SRPA exploration, the TM falls into the energy region of
so-called pygmy dipole resonance (PDR) \cite{Pa07} supposed to
be induced in neutron rich nuclei by a relative translational
oscillation of the neutron skin against the residual N=Z core. So
an interplay of TM and PDR may be expected.  As shown in our
recent study \cite{Rep_PRC_13} within the full
(non-separable) RPA \cite{Rei92},
the PDR energy region indeed embraces
various modes with a strong TM fraction. Moreover, the
vortical flow dominates in the nuclear interior while the irrotational
motion (relevant for E1 transitions in the long-wave
approximation) prevails at the nuclear surface. This point
deserves a further investigation. In particular, a
comparative analysis of experimental data from $(\alpha,\alpha'
\gamma)$ (relevant for both TM and PDR in T=0 channel),
$(\gamma,\gamma')$ observations (e.g. in Sn isotopes
\cite{Endres12} and N=82 isotones \cite{Savran11}), and $(e,e')$
reactions is needed.

In this study, we continue the exploration of the toroidal,
compression and vortical (RW) E1 strengths in various mass regions
using SRPA. While the previous analysis concerned $^{208}$Pb
\cite{Kv11} and Sn isotopes \cite{Sn_PS_13}, the present study
concentrates on Sm isotopes, from spherical $^{144}$Sm to axially
deformed $^{154}$Sm. Thereby we look particularly at the influence of
the nuclear deformation.  Like in \cite{Sn_PS_13}, we use a
representative set of Skyrme forces (SLy6 \cite{Ch97}, SkT6
\cite{To84}, SVbas \cite{Kl09}, SkM* \cite{Ba82}, and SkI3 \cite{Re95})
covering a wide range of the isoscalar effective mass, $m_0/m = 1 -
0.58$.  The isoscalar (T=0), isovector (T=1), and pure proton (elm)
channels of E1 excitations are considered. Like in
\cite{Kv11,Sn_PS_13}, the relative contributions to the strengths of
the convection $j_{\text{con}}$ and magnetization $j_{\text{mag}}$
nuclear currents are inspected.
To demonstrate
the ability of our approach and accuracy of different Skyrme
parametrizations, the supplemented characteristics (binding energies,
photoabsorption cross sections, energy-weighted sum rules)
are considered.

The SRPA method used in the present calculations has been
already successfully applied to description of various kinds of
nuclear excitations (electric
\cite{Kv11,Ne07,Nest_IJMPE_08,Kl08,Kv11b} and magnetic
\cite{Ve09,Ne10,Nest_IJMPE_10_M1} giant resonances, E1 strength near
the particle thresholds \cite{Kv11b,Kv09}, and TM/CM/RW modes
\cite{Kv11,Sn_PS_13}) in both spherical and deformed nuclei and was
shown as an efficient and reliable theoretical tool.

 The paper is organized as follows.  In Sec. II the theoretical
 background and calculation scheme are presented.  In Sec. III,
 the supplemented characteristics are inspected and
 numerical results for TM, CM, and RW strengths are discussed. In
 Sec. IV, the conclusions are done.  In Appendix A, the vortical,
 toroidal, and compression flows and their operators are discussed.
 The HD and RW conceptions of the vorticity on nuclear flow are outlined.
 Appendix B sketches the derivation of the RW equations.
 In Appendix C, a basic information about the SRPA method is given.

\section{Theoretical background and calculation scheme}

The main topic of this paper is the influence of nuclear
deformation on the toroidal, compressional and vortical dipole
strength functions. The corresponding transition operators are \cite{Kv11}
\begin{eqnarray} \label{29}
&& \hat{M}(\rm{tor};\:1\mu) = - \frac{2}{2c\sqrt{3}}\:\int\:d^3r \: \hat{\vec{j}}_{nuc}(\vec{r})
\nonumber \\
&&  \cdot\:\left[\:\frac{\sqrt{2}}{5} \:r^2 \:\vec{Y}_{12\mu}(\hat{r}) +
(r^2 - \delta_{T,0}
\langle r^2\rangle_0)\:\vec{Y}_{10\mu}(\hat{r})\:\right],
\end{eqnarray}
\begin{eqnarray} \label{30}
&& \hat{M'}(\rm{com};\:1\mu) = \frac{1}{10}\:\int\:d^3r \: \hat{\rho}(\vec{r})
\nonumber \\
&& \qquad \qquad \quad \cdot\: \left[\: r^3 - \delta_{T,0} \:\frac{5}{3}\:
\langle r^2 \rangle_0 \:r\:\right]\:Y_{1\:\mu}(\hat{r}),
\end{eqnarray}
\begin{equation} \label{31}
 \hat{M}(\rm{vor};\:1\mu) =
 -\frac{i}{5c}\:\sqrt{\frac{3}{2}}\:\int\:d^3r \:r^2\:
\hat{\vec{j}}_{\text{nuc}}(\vec{r}) \cdot
\vec{Y}_{12\mu}(\hat{r}),
\end{equation}
where $\hat{\vec{j}}_{\rm{nuc}}(\vec{r})$ and
$\hat{\rho}(\vec{r})$ are operator of nuclear current and nuclear
density, respectively. Symbols $Y_{\lambda \mu}(\hat{r})$ and
$\vec{Y}_{\lambda l \mu}(\vec{r})$ stand for spherical harmonics
and vector spherical harmonics, respectively, and $\langle
r^2\rangle_0 = \int\:d^3r \: \rho_0(\vec{r}) \:r^2$ is the
ground state square  radius. The derivation of these operators and
their connection to the long-wavelength limit of the standard E1
operator
\begin{equation}
\hat{M}(E\:1 \mu) = - \int d^3r \:\hat{\rho}(\vec{r}) \: r \: Y_{1\mu}(\hat{r})
\label{E1op}
\end{equation}
can be found in the Appendix A and Ref. \cite{Kv11}.
The toroidal, compressional, and vortical strength functions
\begin{eqnarray} \label{39}
&& S\:'_{\gamma}(E1\:,\;E) =
\nonumber \\
&& \sum_{\mu=0,\mp1} \sum_{\nu} \:
|\:<\nu|\:\hat{M}(\gamma;\;1\mu)\:|0> \:|^2
\:\xi_{\Delta}(E-E_{\nu})
\nonumber \\
\end{eqnarray}
are calculated in the framework of the Skyrme SRPA approach
\cite{Ne02,Ne06}, see Appendix C for more detail. In the above
expression, $\gamma$ labels the TM ,CM, and RW strengths
determined by the operators (\ref{29}), (\ref{30}), and
(\ref{31}), respectively. The indices $\nu$ stand for the RPA states
with the energies $E_{\nu}$, $|0>$ is the RPA ground state.
Further,
\begin{equation}\label{40}
\xi_{\Delta}(E-E_{\nu}) = \frac{1}{2\pi}\:\frac{\Delta}{(E-E_{\nu})^2
 + (\frac{\Delta}{2})^2}
\end{equation}
is the Lorentz weight with the averaging parameter $\Delta$. The
averaging is needed for the convenience of comparison of the
results with the experimental data and to simulate roughly the
smoothing effects beyond the SRPA (escape widths and coupling to
complex configurations). For the broad and poorly known TM,CM, and
RW strengths, a constant averaging width $\Delta$=1 MeV is
optimal. The strengths are analyzed
in T=0, T=1, and electromagnetic ($elm$) channels characterized by
the effective charges $e_q^{\rm{eff}}$ and gyromagnetic ratios
$g_q^{\rm{eff}}$
\begin{eqnarray} \label{44}
T=0: && e_n^{\text{eff}}=e_p^{\text{eff}}=1, \;
g_{n,p}^{\text{eff}}=\frac{\zeta }{2}\:(g_n + g_p) ,
\\
\label{45}
T=1: && e_n^{\text{eff}}=-e_p^{\text{eff}}=-1,
\;g_{n,p}^{\text{eff}}=\frac{\zeta}{2}\:(g_n - g_p) ,
\\
\label{46}
elm: && e_n^{\text{eff}}=0,\:e_p^{\text{eff}}=1, \quad
g_{n,p}^{\text{eff}}=\zeta g_{n,p} \; ,
\end{eqnarray}
where $g_{n} = -3.82$ and $g_{p} = 5.58$ are free neutron and
proton gyromagnetic ratios, $\zeta \approx 0.7$ is the usual
quenching factor \cite{Al89}. See details in Appendix A.

The photoabsorption cross-section is fully determined by the $E1$
transitions and thus reads \cite{Ri80}
\begin{equation} \label{43}
\sigma_{\text{phot}}(E) = \frac{16\:\pi^3\:\alpha_{e}}{9\:e^2} \;E\: S(E1;\:E),
\end{equation}
where $\alpha_{e}=1/137$ is the fine-structure constant and $S(E1;\:E)$
is the strength function calculated with the standard dipole operator
\begin{equation}\label{E1}
 \hat{M} (E1\mu)
  = \frac{N}{A}\sum_{p=1}^Z r_p Y_{1\mu}(\Omega_p)
  -
  \frac{Z}{A}\sum_{n=1}^N r_n Y_{1\mu}(\Omega_n) \; .
\end{equation}

For the photoabsorption cross-section, detailed experimental data are
available \cite{atlas,janis}. So it is worth to compute this observable
more accurately.
It should be taken into account that
i) the escape widths appear above the particle emission thresholds and
grow with energy due to the widening of the emission phase space,
ii) the collisional widths, induced by the coupling with complex
configurations, also increase with the excitation energy.
To simulate these trends,
one should use in (\ref{40}) an energy-dependent averaging parameter
$\Delta (E)$.  This can be done by implementing a double folding
scheme \cite{Kv11b}. We calculate first the strength function
(\ref{39}) with the operator (\ref{E1}) by using a small but fixed
value of $\Delta$. This gives the strength distribution $S'(E1;\;E)$
very closed to one obtained in RPA but with an equidistant energy
grid. This strength is then folded again by using an energy dependent
$\Delta (E)$:
\begin{equation}\label{41}
S(E1;\;E) = \int\:dE'\:S\:'(E1;\;E')\;\xi_{\Delta (E')}(E-E').
\end{equation}
In the present study, we use a simple linear  dependence
\begin{equation}\label{42}
    \Delta (E') = \left\{ \begin{array}{ll}
\Delta_0 & \mbox{for } E'\leq E_{\mathrm{th}} \\
\Delta_0 + a (E' - E_{\mathrm{th}}) & \mbox{for } E' > E_{\mathrm{th}}
\end{array}\right.
\end{equation}
where $E_{\text{th}}$ is the energy of the first emission threshold and
$\Delta_0=0.1$ MeV  is a minimal width.
The parameter $a$ is chosen to reproduce the GDR strength
distribution.

In order to test the sensitivity of the modes, we consider a selection
of sufficiently different Skyrme parametrizations (SkT6 \cite{To84},
SVbas \cite{Kl09}, SkM* \cite{Ba82}, SLy6 \cite{Ch97}, and SkI3
\cite{Re95}) covering a wide range of isoscalar effective
masses: $m_0/m$=1, 0.9, 0.79, 0.69, and 0.58, respectively. All
the parametrizations provide a good description of basic
ground properties of nuclei.

The configuration space in the calculations covers all particle-hole
(two-quasiparticle) states with an excitation energy up to $E_{\rm{cut}} \approx 175$
MeV. Such a big basis allows to exhaust in the SRPA calculations about
100\% of the isovector Thomas-Reiche-Kuhn (TRK) energy-weighted sum rule (EWSR)
\cite{Ri80} for the GDR and isoscalar Harakeh's EWSR for the ISGDR \cite{Ha01}
(see also an extensive discussion at the end of the next subsection
and Appendix C). Besides that, the large basis is crucial to lower the
energy position of the spurious E1(T=0) peak towards the correct zero
energy. In the present study, it lies in spherical $^{144}$Sm at
$1.5-2.0$ MeV, depending on the Skyrme parametrization.  This is not
so good as in some full RPA calculations for $^{208}$Pb, which use a
larger configuration space and manage to yield the spurious peak below
$\sim$ 0.6 MeV, see e.g. SLy5 \cite{Co13} and SLy6 \cite{Rep_PRC_13}
results.  However, the present value is sufficient for our aims since
the T=0 strengths of interest are situated at $E>$ 5 MeV and further
extension of the configuration space does not significantly affect them.
To be on the safe side, the TM, CM, and RW strength functions are given
below only in the range $E>$ 5 MeV.

The calculations exploit a 2D representation in cylindrical
coordinates using a mesh size of 0.3 fm and a calculation box of 21
fm. For open-shell isotopes, we use zero-range pairing forces.  The
Hartree-Fock-Bogoliubov (HFB) equations are implemented at the BCS
level \cite{Ben00}.  For closed shell nuclei, this reduces
automatically to a Hartree-Fock (HF) treatment.

\section{Results and Discussions}

\subsection{Supplemented characterisics}
\begin{figure}[b]
\includegraphics[width=9cm]{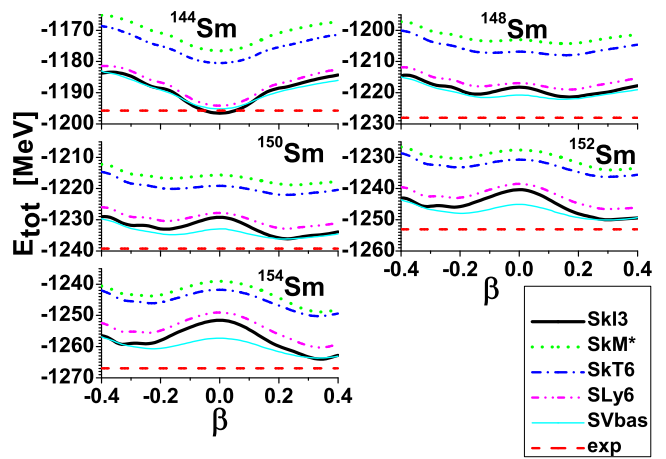}
\vspace{2mm}
\caption{ (color online) HF/HFB binding energies of
Sm isotopes versus the dimensionless parameter of quadrupole deformation
$\beta$, calculated with different Skyrme parametrizations. The
binding energies at an
equilibrium deformation are to be compared with the experimental
values (dashed horizontal lines) obtained from mass measurements
\cite{mass}}.
\end{figure}

\begin{figure}[t]
\includegraphics[width=9cm]{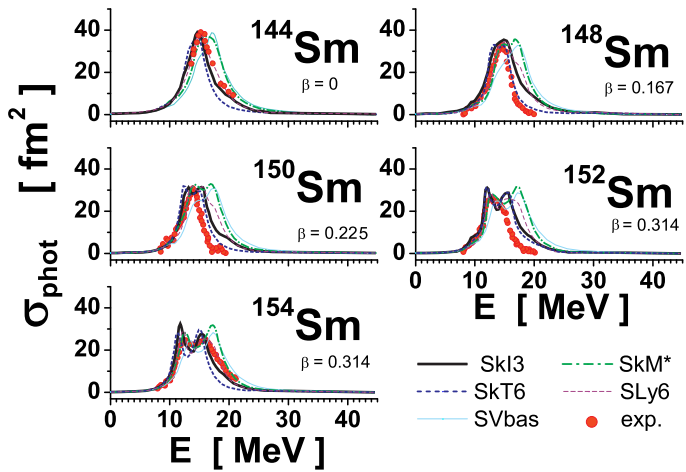}
\vspace{5mm}
\caption{ (color online) Photoabsorption cross section versus
excitation energy in
 $^{144-154}$Sm isotopes, calculated for different
 Skyrme parametrizations and compared
 with experimental values \cite{janis}. For each isotope, we indicate
 the dimensionless quadrupole deformation $\beta$
 of the prolate ground state computed with the force SLy6.}
\end{figure}

In Figures 1 and 2, some basic features (binding energies and photoabsorption
cross sections) are presented to demonstrate the accuracy of our
approach. In addition to spherical $^{144}$Sm and deformed $^{154}$Sm,
the transitional isotopes $^{148,150,152}$Sm are also shown to
illustrate the trends with development of the deformation. The
treatment of these soft isotopes within RPA is known to be
insufficient and has to be amended by the coupling with complex
configurations \cite{En10,Li08}. Nevertheless, we find useful to
present these RPA results to outline the trends.

In Figure 1, the binding energies of Sm isotopes calculated within the
HF/HFB approach are depicted  in dependence on the dimensionless parameter
of the quadrupole deformation $\beta$.  For all Skyrme forces used here, we see the
pronounced main minima at $\beta$=0 in the spherical $^{144}$Sm and at
$\beta$=0.33-0.34 in the prolate deformed $^{154}$Sm.  In the latter
case, the internal quadrupole moment is $Q$=6.4-6.6 b. Both computed
$\beta$ and $Q$ are in a good agreement with the experimental values
$\beta_{\text{exp}}$=0.34 and $Q_{\text{exp}}$=6.6 b \cite{Raman87}.
In the transitional $^{148,150,152}$Sm, the calculations produce two
shallow minima with roughly the same depth, corresponding to prolate
and oblate shapes, respectively.  Despite all the Skyrme
parametrizations are fitted to experimental binding energies of
selected doubly magic or semi-magic nuclei, in our calculations only
the SLy6, SV-bas, and SkI3 forces reproduce the measured binding
energy \cite{mass} in the semi-magic $^{144}$Sm. The two older
parametrizations SkM$^*$ and SkT6 were tuned to doubly-magic nuclei
only and do not perform so well for the rather soft $^{144}$Sm.

In Figure 2, the calculated photoabsorption cross section is
inspected.  We get a good agreement with experiment \cite{janis} for
most Skyrme forces in the spherical $^{144}$Sm and the well deformed
$^{154}$Sm.  The deformation splitting of the GDR in $^{154}$Sm is
also reproduced. For the transitional isotopes, the agreement is
acceptable as well. In all the nuclei, the main deviation arises
at the high-energy wing of the GDR. This is probably an effect of
neglecting the complex configurations.
\begin{figure*}
\includegraphics[width=12cm]{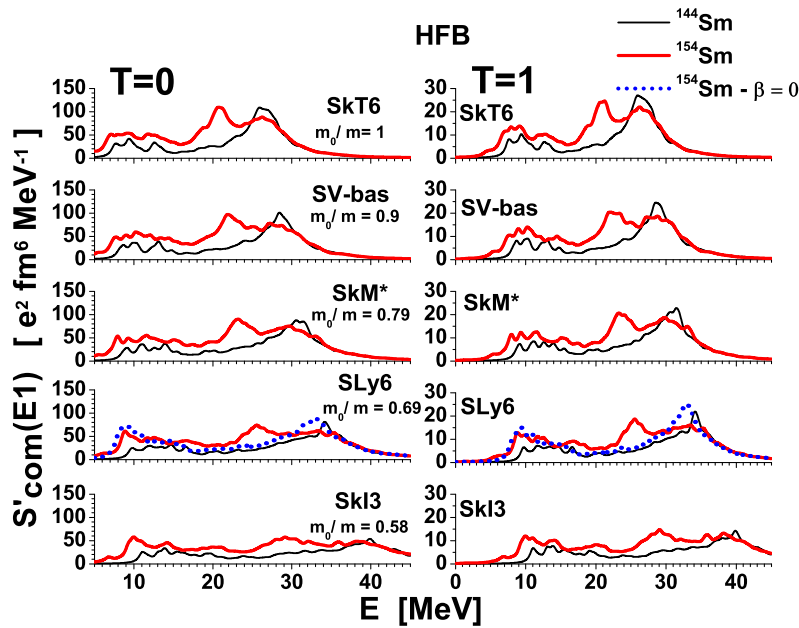}
\caption{ (color online) Pure HF/HFB strength functions for T=0 (left) and
T=1 (right) compression mode in spherical $^{144}$Sm (black thin curve)
and deformed $^{154}$Sm (red bold curve) isotopes, computed with different
Skyrme forces. For SLy6 in $^{154}$Sm, we show also
the strength computed for an artificial ground state forced to
have $\beta$=0, i.e. stay spherical
(blue dashed curve). On the left, the isoscalar effective
masses $m_0/m$ of the forces are listed.}
\end{figure*}

Additional useful insight may be obtained from the analysis of the sum
rules. As was mentioned above, the TRK sum rule for the
photoabsorption cross section is exhausted in our calculations by
about 100$\%$, see details in Appendix C.
For the purposes of the present study, it is crucial to check the
sum rule for the ISGDR \cite{Ha01}
\begin{equation}\label{Har_EWSR}
\text{EWSR}_{\text{ISGDR}} = \frac{1}{100} \left[ \frac{3
\hbar^2}{8 \pi m} \:A\:( 11 \left<r^4\right>_0 - \frac{25}{3}
\left<r^2\right>_0^2 ) \right] ,
\end{equation}
obtained for the transition operator (\ref{30}) and
$\left<r^4\right>_0=\int\:d^3r \: \rho_0(\vec{r}) \:r^4$. The results
are shown in Table \ref{tab1} and they prove that this sum rule
is also nicely fulfilled. Thus exhausting both isovector TRK and and
isoscalar ISGDR E1 sum rules confirms that our configuration space is
sufficiently large.
\begin{table}
\begin{center}
\caption{\label{tab1} The ISGDR (\ref{Har_EWSR}) and RPA EWSR
(in $\mathrm{e^2\:fm^6} 10^3$ MeV)
for CM(T=0) in $^{144,154}$Sm, computed with different Skyrme
forces.}
\begin{tabular}{|c|c|c|c|c|c|c|}
\hline
\multicolumn{2}{|r|}{ } & SkI3 & SLy6 & SkM* & SVbas & SkT6 \\
\hline
\multicolumn{2}{|r|} {ISGDR}
& 24.4 & 24.7 & 25.1 & 24.3 & 24.4 \\
\multicolumn{2}{|r|}{ $^{144}$Sm \qquad RPA
} & 25.0 & 25.4 & 24.8 & 23.1 & 24.3
\\
\hline \hline \multicolumn{2}{|r|}{ ISGDR}
& 34.1 & 33.8 & 34.1 & 33.2 & 33.2 \\
\multicolumn{2}{|r|}{ $^{154}$Sm \qquad RPA
} & 34.8 & 34.0 & 33.5 & 32.6 & 33.1
\\
\hline
\end{tabular}
\end{center}
\end{table}
\begin{figure}
\includegraphics[width=8cm]{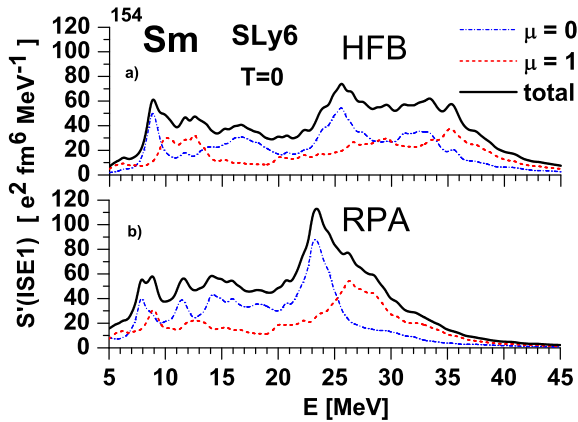}
\caption{ (color online) Total (black bold), $\mu=0$ (red dash),
and $\mu=1$ (blue dash-dotted) CM(T=0) strength functions in the
deformed nucleus $^{154}$Sm, computed with the force SLy6 within
HFB (upper panel) and RPA (bottom panel) approaches.}
\end{figure}
\begin{figure*}
\includegraphics[width=12cm]{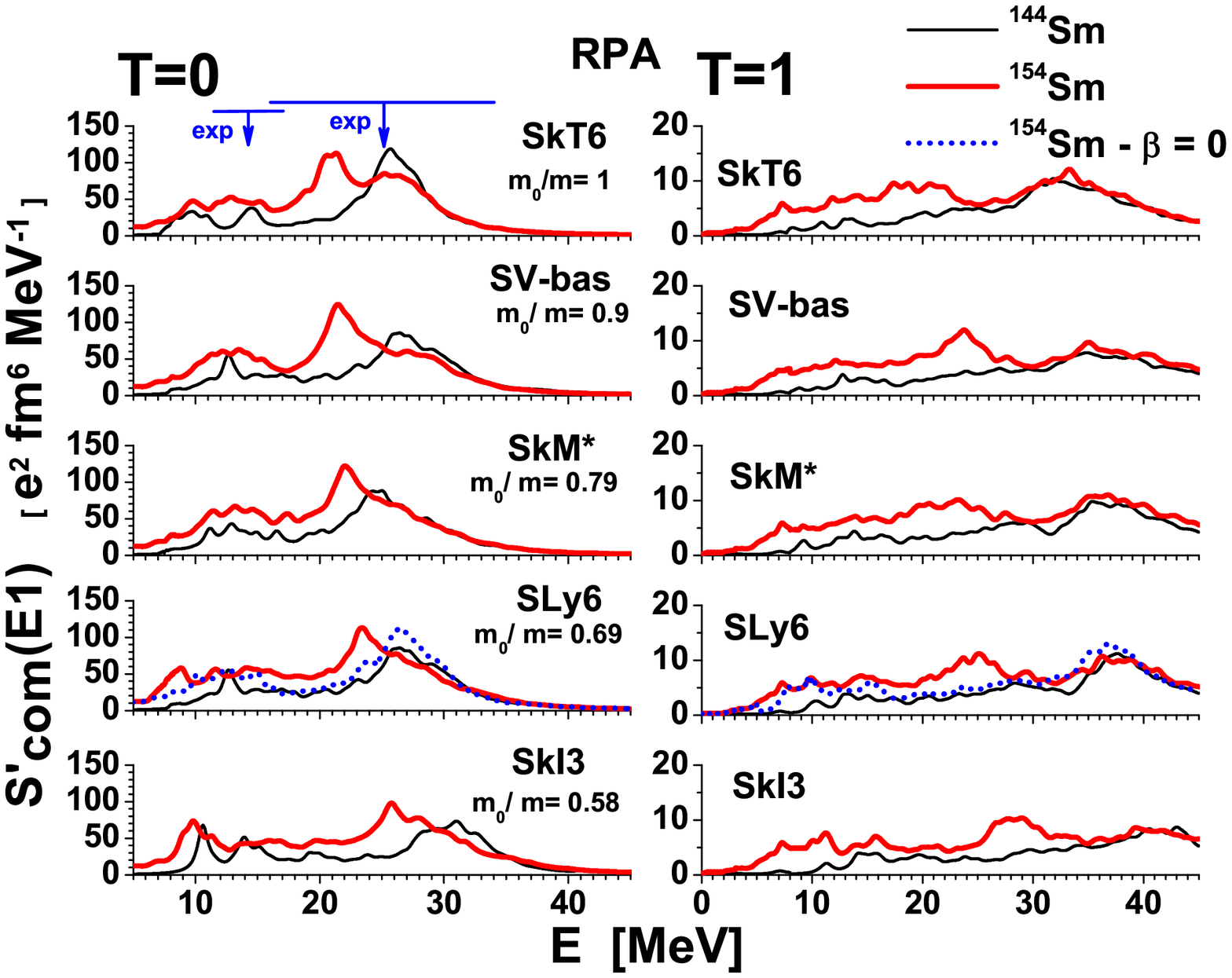}
\caption{ (color online) The same as in Fig. 3 but for RPA
compressional (CM) strength functions. The widths and energy
centroids of the low- and high-energy ISGDR branches observed in
($\alpha, \alpha'$) reaction \protect\cite{Itoh_PRC_03,Uchi04} are
denoted at the upper/left panel.}
\end{figure*}
Altogether, the above results demonstrate reliability of our approach
in description of static and E1 dynamical properties of spherical and
deformed nuclei and thus justifies its further application to
TM/CM/RW E1 strength functions.

\subsection{TM, CM, and RW strength functions}

In Figure 3, the E1(T=0) and E1(T=1) strength functions for the CM in
spherical $^{144}$Sm and deformed $^{154}$Sm are shown for the chosen
set of Skyrme forces. The strengths are computed purely within the
HF/HFB approach (without the residual interaction) by using the
transition operator (\ref{30}).  As seen from the figure, the
calculations give two broad CM bumps, a smaller one at low-energy
and a large one at high-energy, which is in accordance to
previous theoretical and experimental studies, see review
\cite{Pa07}. In $^{154}$Sm, the strength is larger and more uniform
than in $^{144}$Sm, as expected for a heavier and more deformed
nucleus.

\begin{figure*}[t]
\includegraphics[width=12cm]{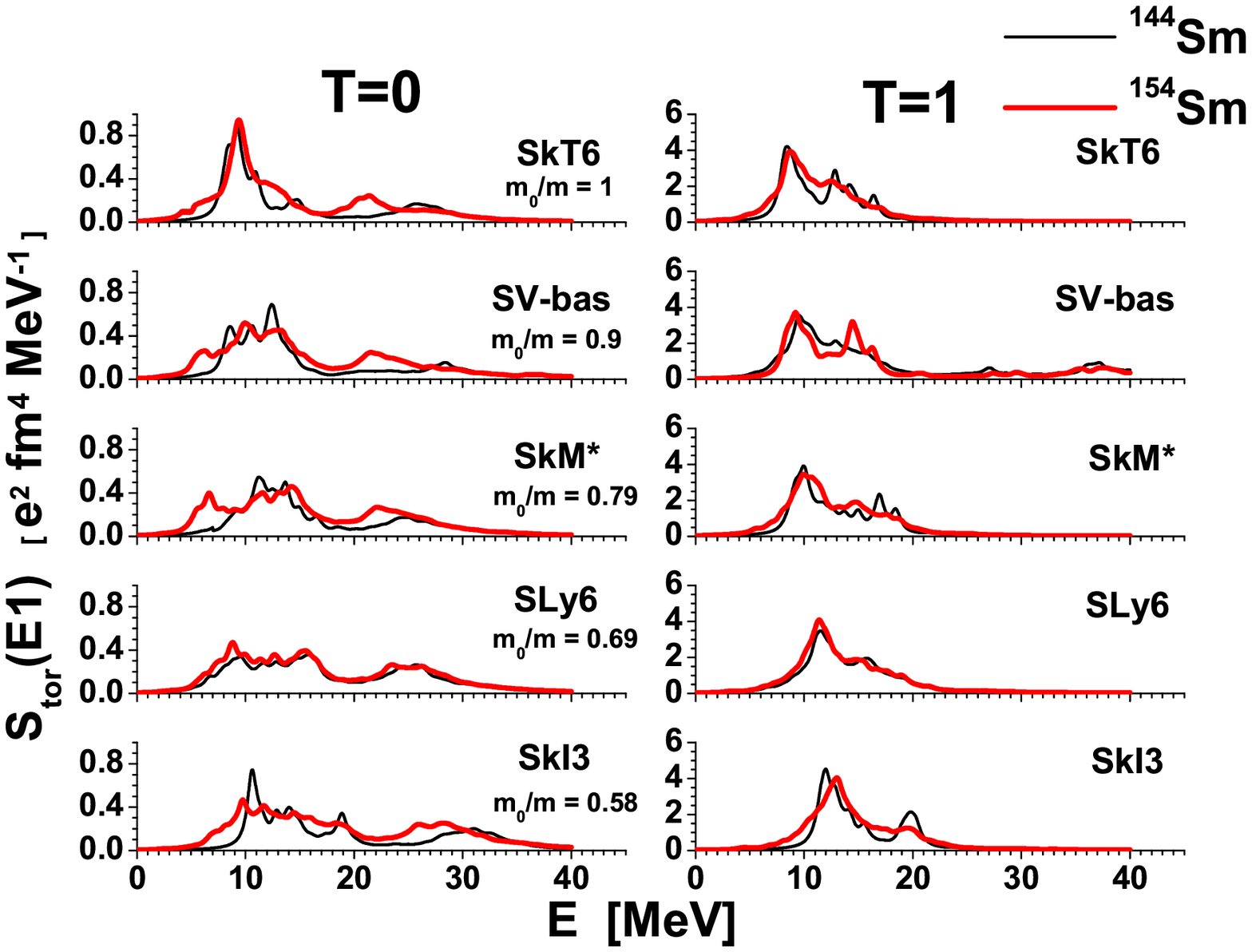}
\caption{ (color online) The same as in Fig. 3 but for the RPA
toroidal strength function.}
\end{figure*}
\begin{figure*}
\includegraphics[width=12cm]{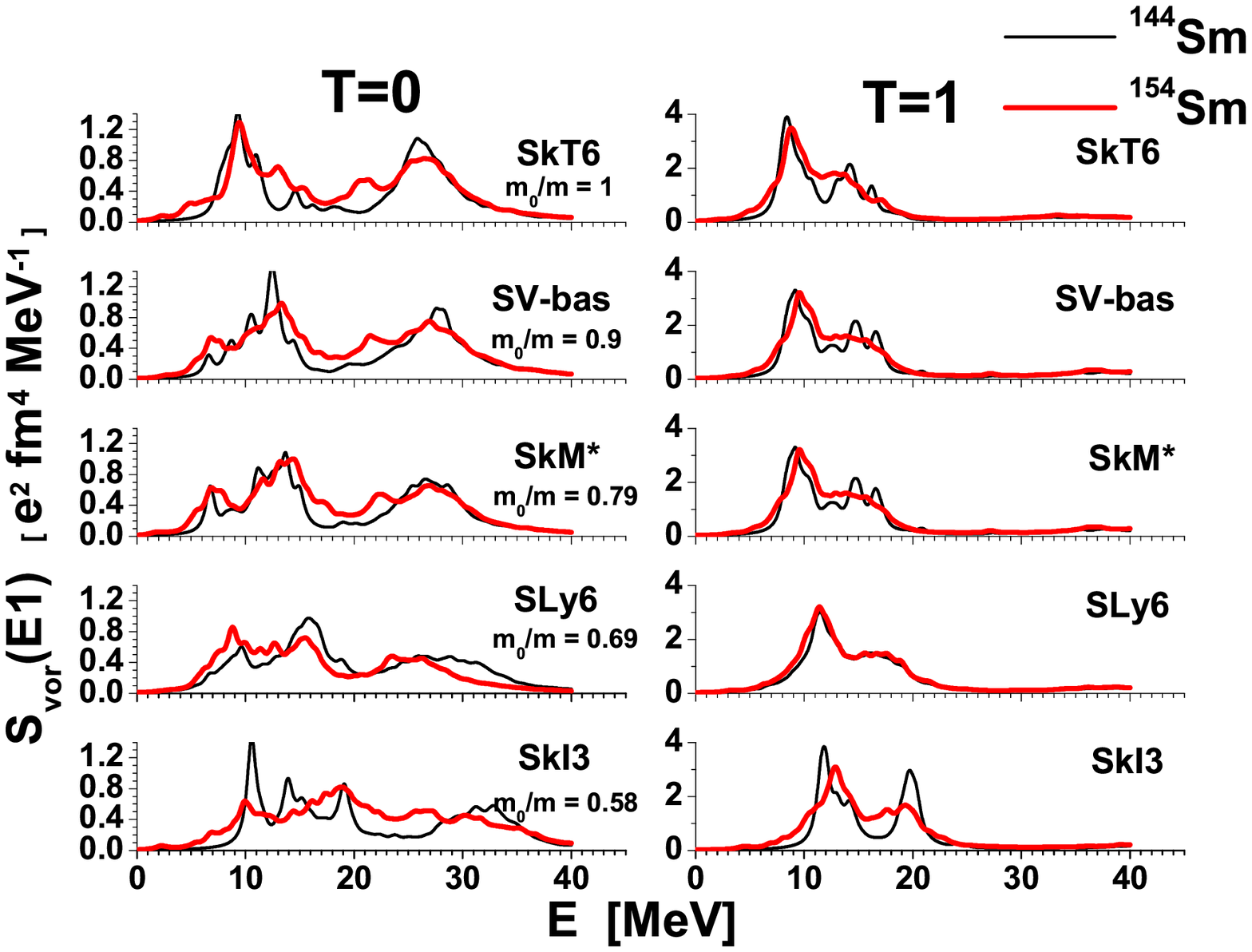}
\caption{ (color online) The same as in Fig. 3 but for the RPA
vortical (RW) strength function.}
\end{figure*}

Three non-trivial effects are visible in both T=0 and T=1 channels.
First, in $^{154}$Sm we see a substantial growth of the CM strength in
the region $4\leq E \leq 11$ MeV where the PDR is supposed to
appear. The comparison with the constrained case $\beta$=0 (no
deformation) for the force SLy6 indicates that this effect should be
mainly ascribed
to increasing the number of neutrons (from magic 82 to 92). Indeed,
unlike $^{144}$Sm, in $^{154}$Sm particular low-energy
s.p. transitions $\nu\nu [1h_{9/2} \leftrightarrow 2f_{7/2}]$ become
active, which may lead to the growth of the strength. However, following
Fig. 4, where the deformation splitting of $\mu$=0 and
$\mu$=1 CM(T=0) components for SkT6 and SLy6 forces is demonstrated,
the deformation effect is also important. Just due to deformation, the
low-energy and high-energy parts of the region $4\leq E \leq 11$ MeV
are dominated by $\mu$=0 and $\mu$=1 branches, respectively.
Moreover, the comparison of T=0 and T=1 strengths allows to state that
the PDR region may be also separated into two isospin sectors, a
low-energy T=0 and a high-energy T=1 ones. This conjecture finds
support in recent experimental analysis \cite{Endres12,Savran11},
which have split the low-lying E1 spectra into an upper and lower
energy sectors.  The upper sector is excited solely in
$(\gamma,\gamma')$ and, therefore, has T=1 nature.  The lower sector,
instead, is composed of levels excited in both $(\gamma,\gamma')$ and
($\alpha,\alpha' \gamma)$ and thus may be related to T=0 excitations.

Second, Fig. 3 shows that in $^{154}$Sm a substantial bump appears at
20-30 MeV. This is particularly pronounced in T=0. The comparison to
the case $\beta$=0 indicates that the bump is caused by the
deformation.  Fig. 4a) shows that this is just the $\mu$=0 branch of
the high-energy CM. We also see in Fig. 4a) a huge deformation
splitting ($\sim$ 10 MeV) of the CM(T=0) strength computed with SLy6
force (similar results are obtained for other Skyrme forces). Thus
deformation gives for the high-energy CM indeed dramatic effect.

Third, Fig. 3 shows that the CM strength in general and low-energy and
high-energy CM bumps in particular are noticeably upshifted with
decreasing $m_0/m$. This takes place for both isotopes $^{144}$Sm and
$^{154}$Sm and in both channels, T=0 and T=1. The effect is
straightforwardly explained by well known spread of s.p.  spectra
below the Fermi level with decreasing $m_0/m$, see examples in
\cite{Nest_PRC_04}.
\begin{figure*}[t]
\includegraphics[width=12cm]{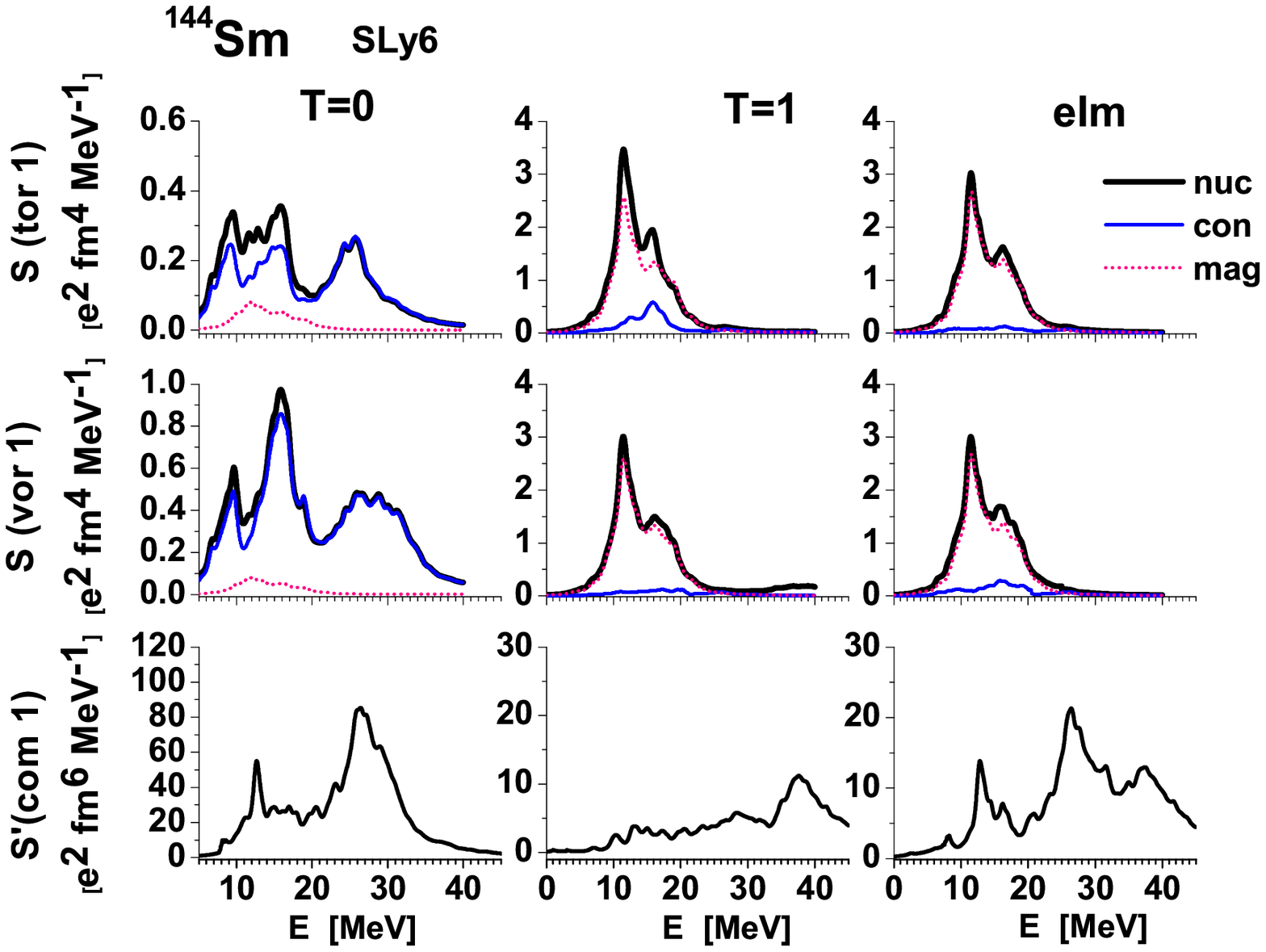}
\caption{ (color online) RPA toroidal (upper), RW vortical
(middle), and compression (bottom) strength functions calculated
in T=0 (left), T=1 (middle), and electromagnetic (right) channels
with SLy6 Skyrme force in spherical $^{144}$Sm. For TM and RW, the
contributions from  total (black/bold line),  convection
(blue/thin line), and magnetization (red/dotted line) currents are
depicted.}
\end{figure*}
\begin{figure*}[t]
\includegraphics[width=12cm]{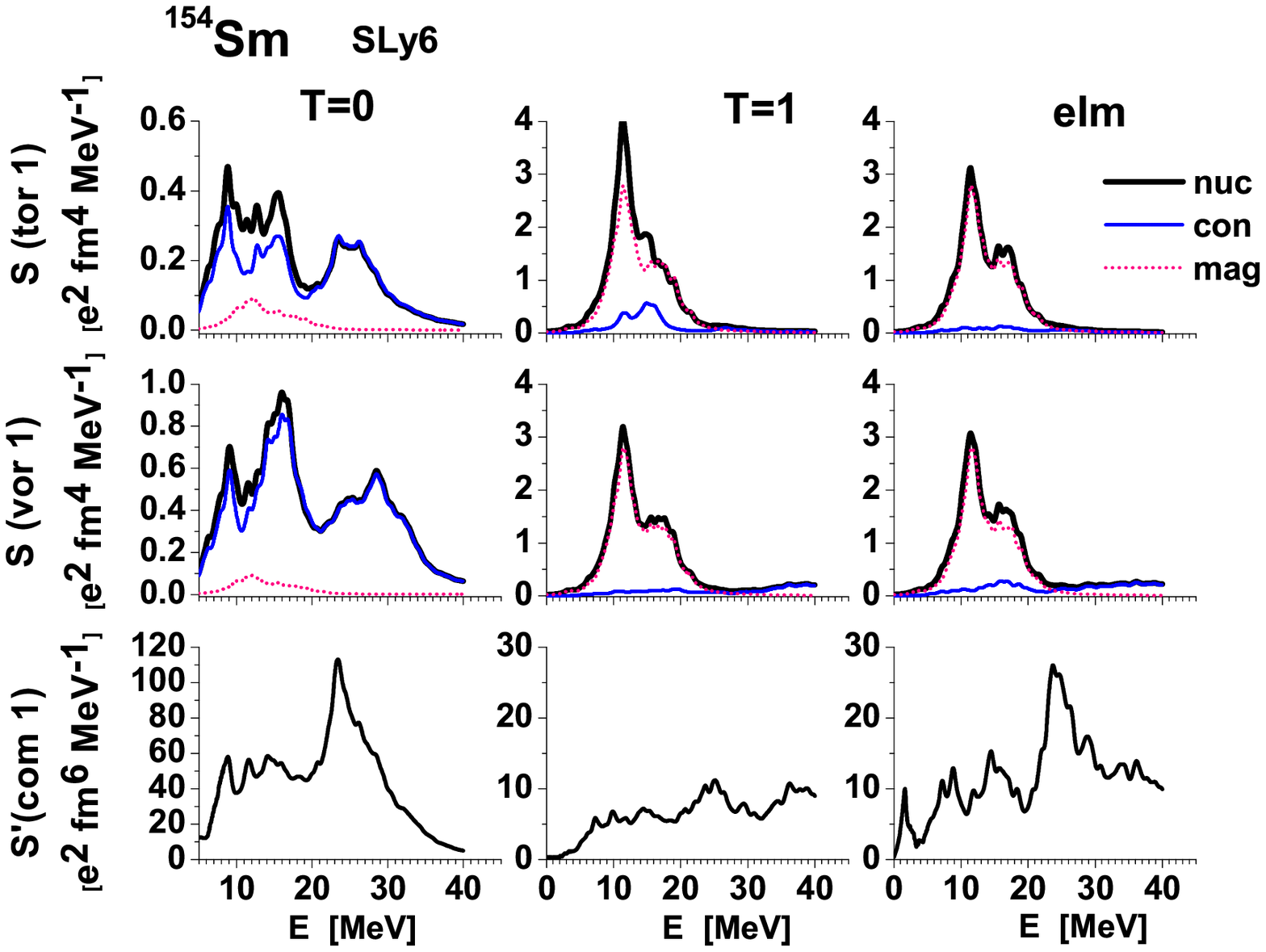}
\caption{ (color online) The same as in Fig.8 but for deformed
$^{154}$Sm.}
\end{figure*}

In Figure 5, the RPA CM strength functions are depicted.  The residual
interaction significantly downshifts (upshifts) the strength in T=0
(T=1) channels. Nevertheless all three effects discussed above for
HF/HFB case remain the same. The $(\alpha,\alpha')$ experiment
\cite{Itoh_PRC_03,Uchi04} gives for the ISGDR in $^{144}$Sm two bumps
with the energies and widths
E=14.2 MeV, $\Gamma$=4.8 MeV and E=25.0 MeV, $\Gamma$=19.9 MeV. The
narrow low-energy and broad high-energy bumps of the ISGDR are
commonly treated as TM \cite{Vret02,Kv03,Mis06} and CM
\cite{Vret00,Colo00}, respectively. Alternatively, both bumps may be
treated as merely CM branches with the strong CM/TM coupling at a low
energy \cite{Rep_PRC_13}. Following Fig. 5, our calculations for
$^{144}$Sm roughly reproduce the experimental data
\cite{Itoh_PRC_03,Uchi04} for the forces SkT6, SVbas, SkM*, and
SLy6. For SkI3 with a low effective mass $m_0/m$=0.58, the agreement
for the high-energy CM bump is not so good. It seems that Skyrme
forces with a large $m_0/m$ are more preferable for description of
CM.

Note that, following Fig. 4, the residual interaction considerably
decreases the deformation splitting. As compared to the HFB case,
it is reduced for the high-energy CM(T=0) bump from $\sim$ 10 MeV
(HFB) to $\sim$ 5 MeV (RPA). Nevertheless, the deformation
splitting remains very large. Such strong splitting is known only
for the isovector GDR.

In Fig. 6, the RPA TM strength in $^{144,154}$Sm is depicted in T=0
and T=1 channels. Unlike the CM, the TM is concentrated at low energy
while the high-energy strength is rather weak. Following our analysis,
this is because of the instructive (destructive) summation of $\lambda
l$=10 and 12 terms in the TM operator (\ref{29}) in the low-energy
(high-energy) regions. For the CM operator (\ref{26}) (which may be
decomposed in similar manner \cite{Kv11}), we have the opposite
result.

The TM(T=0) strength in Fig. 6 demonstrates the same three effects
discussed above for CM: i) pumping the strength to the PDR region due
to neutron excess/deformation impact, ii) appearance of an appreciable
$\mu$ =0 bump due to the deformation splitting of the high-energy
strength, iii) general upshift of the strength with $m_0/m$. The
manifestation of the effects for TM is weaker than for CM and
noticeably depends on the force: the effects are significant for SkT6,
SVbas, SkM*, SkI3 but small for SLy6. In the T=1 channel, the effects
are much weaker, perhaps because the main player, the high-energy TM,
is almost absent.

In Fig. 7, a similar behavior is observed for the vortical RW
strength generated by the operator (\ref{31}). The main difference from
TM is that low-energy and high-energy RW bumps in T=0 channel are
of a comparable strength. Anyway both TM and RW are strong at 5-15
MeV where the maximal nuclear vorticity is expected \cite{Kv11}.

Figs. 8 compares in $^{144}$Sm the CM, TM, and RW strengths in T=0 and
T=1 channels with the strengths in the electromagnetic ('elm') channel
relevant for $(e,e')$ reaction, where only the proton part of the
transition operators is active, see (\ref{46}).  Fig. 9 does the same
for $^{154}$Sm. It is seen that the strengths in the 'elm' channel are
very similar to T=1 ones. Further Figs. 8 and 9 demonstrate the
contributions of the convection $j_{\rm{con}}$ and magnetization $j_{\rm{mag}}$
nuclear currents (\ref{11}) to the strength. In accordance to the
previous calculations for $^{208}$Pb \cite{Kv11}, the $j_{\rm{con}}$
dominates at T=0 and $j_{\rm{mag}}$ at T=1 (and 'elm').

Finally we remark that for a detailed study it is desirable to go
beyond RPA and take into account more complex configurations. The
low-lying E1 modes, in fact, may couple to two-phonon states
describing collective-core excitations coupled to collective
surface  vibrations, like  an octupole-quadrupole excitation.
Several RPA extensions are already available. Among them, there
are the QRPA plus phonon-coupling model \cite{colo94,CoBo01} and
the separable RPA plus two-phonon approach \cite{NVG_98,Sever_08,Ars}
with Skyrme forces, and the relativistic time-blocking
approximation (RTBA) \cite{Li08} embracing  the anharmonicity by
coupling  the RPA phonons to particle-hole + phonon states.
Alternatively, one may use
the quasiparticle-phonon model (QPM)
\cite{Sol92,Pon96,LoJoP12} with Woods-Saxon separable
potential  and  the equation-of-motion phonon method (EMPM)
with realistic interactions \cite{bianco12}. All these RPA
extensions have been adopted with fair success to describe the
low-energy E1 response in nuclei like $^{208}$Pb (see e.g.
\cite{Tamii11,Polto,bianco12a}) and confirmed an important role
played by the multi-phonon states in E1 low-lying spectra.
However, before performing more detailed investigations with including
multiphonon states, a first exploration with mere RPA exploration is
desirable and this is just what we have presented here.

\section{Conclusion}

The E1 compression, toroidal, and vortical \cite{Ra87} strengths in
even Sm isotopes (from spherical $^{144}$Sm to deformed $^{154}$Sm)
were investigated within the self-consistent separable random-phase
approximation (SRPA) approach \cite{Ne02,Ne06} with Skyrme forces.  A
representative set of different Skyrme parametrizations (SkT6
\cite{To84}, SVbas \cite{Kl09}, SkM* \cite{Ba82}, SLy6 \cite{Ch97},
and SkI3 \cite{Re95}) was used. Three reaction channels were inspected:
isoscalar (T=0), isovector (T=1) and electromagnetic (with the proton
part of the transition operator). Particular attention was paid to
effects of the quadrupole deformation.

As a first step, the accuracy of the scheme was checked with respect
to binding energies, photoabsorption cross sections, and E1
energy-weighted sum rules EWSR (isovector for the GDR and isoscalar
for the ISGDR).
It was shown that the configuration space used in our
calculations is large enough to guarantee a good reproduction of
the sum rules.

We have analyzed in detail the E1 compressional, toroidal, and
vortical strength functions for the chosen variety of the Skyrme
forces. The strengths in the electromagnetic channel (relevant for the
reaction of inelastic electron scattering) are very similar to T=1
strength. In accordance to our previous explorations for $^{208}$Pb
\cite{Kv11} and Sn isotopes \cite{Sn_PS_13}, we find that the
convection nuclear current $j_{\text{con}}$ plays a dominate role in
all T=0 strengths while the magnetization nuclear current
$j_{\text{mag}}$ is crucial in T=1 and electromagnetic channels.

All strength distributions are upshifted to higher energies with
decreasing effective mass $m_0$ which may be related to the
corresponding spreading the single-particle spectra
\cite{Nest_PRC_04}.  Following our analysis, Skyrme forces with a high
affective masses, $m_0/m=$ 0.8--1, are more suitable for the
description of the experimental data for the ISGDR
\cite{Itoh_PRC_03,Uchi04}, whose branches are commonly related to the
toroidal and compression modes (TM and CM).

The most interesting results are obtained for the deformation impact
on the strengths.  There is a dramatic general
redistribution of the compressional strength with deformation when
passing from spherical $^{144}$Sm to deformed $^{154}$Sm. In the
quasiparticle (HFB) approximation, the deformation splitting of the
high-energy CM amounts to as much as 10 MeV. The residual interaction
activated in RPA somewhat decreases the effect and reduces the
splitting to 5 MeV.  However even such a splitting is huge. As a
result, the high-energy CM demonstrates a clear separation into
$\mu$=0 and $\mu$=1 branches. The $\mu$=0 branch is considerably
downshifted and looks more like an individual resonance. This effect
is strong for CM and weaker for other strengths. Besides, it is
stronger in the T=0 channel than in the T=1 channel. It would be
interesting to check our findings in $(\alpha,\alpha')$ experiment for
$^{154}$Sm.

Besides that, the move from $^{144}$Sm to $^{154}$Sm causes a
considerable redistribution of the strength in the low-energy (PDR)
region.  This may be attributed to both increasing the number of
neutrons and deformation. Following our calculations, the PDR region
covers both CM and TM strengths in both T=0 and T=1 channels, see also
discussion \cite{Sn_PS_13,Rep_PRC_13}. Moreover, in accordance to
recent experimental analysis \cite{Endres12,Savran11}, the PDR region
may be separated into T=0 and T=1 sectors. However, a proper
exploration of this detail requires a more involved theoretical
framework taking into account the coupling with complex
configurations.

\section*{Acknowledgments}

The work was partly supported by the GSI-F+E-2010-12,
Heisenberg-Landau (Germany - BLTP JINR), and Votruba - Blokhintsev
(Czech Republic - BLTP JINR) grants. W.K. and P.-G.R. are grateful
for the BMBF support under contracts 05P12ODDUE and support from the
F+E program of the Gesellschaft f\"ur Schwerionenforschung (GSI). The
support of the research plan MSM 0021620859 (Ministry of Education
of the Czech Republic) and the Czech Science Foundation project No
P203-13-07117S to this work are also appreciated.

\appendix

\section{Vorticity conceptions and the relevant transition operators}

\subsection{Velocity and nuclear current fields}

In hydrodynamics (HD) \cite{La87}, the vortical motion of a  fluid
is characterized by a non-vanishing value of the curl of the
velocity field
 \begin{equation}\label{1}
 \vec{\nabla} \times \vec{v}_{\rm{HD}} (\vec{r})  \neq 0
\end{equation}
where  $\vec{v} (\vec{r})$ is defined as a time derivative of the
displacement field, see e.g. \cite{Mis06}.  Instead, a
quantum-mechanical treatment of the nuclear vorticity should
involve the actual nuclear physical observables, namely the
nuclear current defined by the operator
\begin{eqnarray}\label{4}
\hat{\vec{j}}_{\text{nuc}} (\vec{r}) = \sum_{\nu \nu'} \langle\nu|\:\hat{\vec{j}}
(\vec{r})\:|\nu'\rangle \:a_{\nu}^+ a_{\nu'} ,
\end{eqnarray}
where $a_{\nu}^+$, $a_{\nu}$ are single particle (s.p.) creation
and annihilation operators, respectively. The nuclear current
consists of a convection  and a magnetization parts:
\begin{equation}\label{11}
\hat{\vec{j}}_{\text{nuc}} = \hat{\vec{j}}_{\text{con}}+\hat{\vec{j}}_{\text{mag}} .
\end{equation}
Thus the current transition density (CTD) reads
\begin{equation}\label{3}
\langle\nu|\:\hat{\vec{j}}_{\text{nuc}} (\vec{r})\:|\nu'\rangle =
\langle\nu|\:\hat{\vec{j}}_{\text{con}}(\vec{r})\:|\nu'\rangle +
\langle\nu|\:\hat{\vec{j}}_{\text{mag}} (\vec{r})\:|\nu'\rangle.
\end{equation}
The convection CTD  is given by
\begin{eqnarray}\label{4a}
\langle\nu|\:\hat{\vec{j}} (\vec{r})_{\text{con}}\:|\nu'\rangle
=\frac{i\:e_{q}^{\text{eff}} \hbar}{2 m} \: \nonumber\\
\times \left[
(\vec{\nabla} \Psi_{\nu}(\vec{r}))^+ \Psi_{\nu'}(\vec{r})
- \Psi_{\nu}^+(\vec{r}) \vec{\nabla} \Psi_{\nu'}(\vec{r}) \right],
\end{eqnarray}
where $\Psi_{\nu}^+(\vec{r})$ are s.p. wave functions and
$e_{q}^{\text{eff}}$  is a proton (q=p)  or neutron (q=n)
 effective charge. The magnetization CTD has the form
\begin{equation}\label{7}
\langle\nu|\:\hat{\vec{j}}_{\text{mag}}(\vec{r})\:|\nu'\rangle =
 \frac{e \hbar}{2m} \: g_{q}^{\text{eff}} \:\Psi_{\nu}^+(\vec{r})
\: (\vec{\nabla} \times \vec{\sigma})\: \Psi_{\nu'}(\vec{r}),
\end{equation}
where $g_{q}^{\text{eff}}$ is the effective gyromagnetic ratio.

As shown below, the nuclear vorticity may be defined either following a
semi-classical HD approach, see e.g. \cite{La87,Mis06}, or a
quantum-mechanical RW route \cite{Ra87}.

\subsection{HD treatment of the vorticity}

In this approach,   the spin part of the nuclear current is usually neglected.
Then one may define the velocity operator $\hat{\vec{v}}_{\rm{HD}}$ through
the nuclear current operator as
\begin{eqnarray}\label{8}
\hat{\vec{j}}_{\text{nuc}}(\vec{r}) = \rho_{0}({\vec{r}}) \: \hat{\vec{v}}_{\rm{HD}}(\vec{r}),
\end{eqnarray}
where $\rho_{0}({\vec{r}})$ is the  density in the ground state.
After taking the curl of the current (\ref{8}), we obtain the quantity
\begin{eqnarray}\label{10}
&& \rho_{0}({\vec{r}}) \: \vec{\nabla} \times \hat{\vec{v}}_{\rm{HD}}(\vec{r}) =
\vec{\nabla} \times \hat{\vec{j}}_{\text{nuc}}(\vec{r}) -
\vec{\nabla} \rho_{0}({\vec{r}}) \times \hat{\vec{v}}_{\rm{HD}}(\vec{r}) =
\nonumber \\
&& = \vec{\nabla} \times \hat{\vec{j}}_{\text{nuc}}(\vec{r}) - \vec{\nabla} \rho_{0}({\vec{r}}) \times
\frac{\hat{\vec{j}}_{\text{nuc}}(\vec{r})}{\rho_{0}({\vec{r}})}
\end{eqnarray}
which may be adopted to measure  the nuclear vorticity.
This definition, however, does not take into account that
the charge  and current densities are not independent but
related by the continuity equation.

\subsection{RW treatment of the vorticity}

In order to eliminate the drawback of the hydrodynamical definition, the RW prescription
\cite{Ra87} starts with decomposing  the nuclear current into an irrotational and
a vortical components
\begin{equation}\label{11}
\hat{\vec{j}}_{\text{nuc}}(\vec{r}) = \hat{\vec{j}}_{\text{irrot}}(\vec{r})
 + \hat{\vec{j}}_{\text{\text{vort}}}(\vec{r})
\end{equation}
It is further assumed that the quantity
\begin{eqnarray}\label{12}
 \hat{\vec{w}}(\vec{r}) &\equiv& \vec{\nabla} \times \hat{\vec{j}}_{\rm{vort}}(\vec{r}) =
\nonumber \\
&=& \vec{\nabla} \times \hat{\vec{j}}_{\text{nuc}}(\vec{r}) - \vec{\nabla}
\times \hat{\vec{j}}_{\text{irrot}}(\vec{r})
\end{eqnarray}
is a measure of a vortical flow. This motion sets in once
$\langle f|\:\hat{\vec{w}}(\vec{r})\:|i\rangle \neq 0$ and vice versa.

One proceeds with a multipole decomposition in terms of
spherical harmonics  $Y_{\lambda \mu}(\hat{r})$ or vector spherical
$\vec{Y}_{\lambda l \mu}(\hat{r})$ harmonics obtaining
\begin{equation}\label{15}
\langle f|\:\hat{\vec{w}}_{\rm{nuc}}(\vec{r})\:|i\rangle = \sum_{l \lambda \mu}
\: b^{(fi)}_{\lambda \mu} \: w_{\lambda l}(r) \;
\vec{Y}_{\lambda l \mu}(\hat{r}),
\end{equation}
where
\begin{equation}
b^{(fi)}_{\lambda \mu} = \frac{(j_i m_i \lambda \mu | j_f m_f )}{\sqrt{2 j_f +1}}
\nonumber
\end{equation}
and
\begin{equation}\label{18}
w_{\lambda \lambda}^{(fi)}(r) = \sqrt{\frac{2 \lambda+1}{\lambda}} \:
\left(\frac{d}{dr} + \frac{\lambda+2}{r} \right)\; j^{(fi)}_{\lambda \: \lambda+1}(r),
\end{equation}
which follows from applying the   continuity equation.

Using the the multipole decomposition, the second term in (\ref{12}) can
be written as
\begin{equation}\label{19}
\left[ \vec{\nabla} \times \hat{\vec{j}}_{\rm{irrot}}(\vec{r}) \right]_{\lambda} =
\frac{1}{\lambda} \; \left[ ( \vec{\nabla} \hat{\rho}(\vec{r}))
\times \hat{\dot{r}} \right]_{\lambda}.
\end{equation}
This quantity  resembles  the second term in the hydrodynamic
equation (\ref{10}). It is therefore legitimate to  adopt   the
matrix element (\ref{15})    as a measure of vorticity \cite{Ra87}.

\subsection{RW vortical operator}

The  vortical multipole operator was introduced
in   \cite{Kv11}. One starts with the  standard $E\lambda$ multipole operator  \cite{Bo69}
\begin{eqnarray}
   \label{20}
&& \hat{M}(E\lambda\mu, k)
= \frac{(2 \lambda + 1)!!}{c k^{\lambda+1}}
\sqrt{\frac{\lambda}{\lambda+1}}
\\
&& \qquad \!\cdot\int d^3r \: [\: j_{\lambda}(kr)\:
\vec{Y}_{\lambda \lambda \mu}(\hat{\vec r}) \:]
\cdot [ \vec{\nabla} \times \hat{\vec{j}}_{\text nuc}(\vec{r})].
\nonumber
\end{eqnarray}
After expanding the  spherical Bessel functions $j_{\lambda}(kr)$ in powers of $(kr)$
and using the continuity equation, one  obtains \cite{Kv03}
\begin{equation}\label{22}
\hat{M}(E \:\lambda \mu;\:k) = \hat{M}(E \lambda \mu)
+ k\;\hat{M}(\rm{tor} ;\: \lambda \mu) +\ldots,
\end{equation}
where $\hat{M}(E \lambda \mu)$ and $\hat{M}(\rm{tor} ;\: \lambda \mu)$ are,
respectively, the long-wavelength limits of the
standard  and   toroidal $E\lambda$ multipole operators \cite{Kv03,Kv11}
\begin{eqnarray} \label{23}
&& \hat{M}(E \:\lambda \mu) = - \int \: d^3r \:\hat{\rho}(\vec{r})
\:r^{\lambda} \: Y_{\lambda \mu}(\hat{r}),
\\
&& \hat{M}(\rm{tor};\: \lambda \mu) = - \frac{i}{2c}\:\sqrt{\frac{\lambda}{2\lambda+1}}
\int \: d^3r \hat{\vec{j}}_{nuc}(\vec{r})
 \\
&& \quad \cdot r^{\lambda+1} \left[ \:\vec{Y}_{\lambda \:\lambda-1\:\mu}(\vec{r})
 + \sqrt{\frac{\lambda}{\lambda+1}}
\:\frac{2}{2\lambda+3} \:\vec{Y}_{\lambda \:\lambda+1\:\mu}(\vec{r}) \:\right].
\nonumber
\end{eqnarray}
The minus sign in  $\hat{M}(E \:\lambda \mu)$  is appropriate for   excitations processes (  $E_f > E_i$).

According to the RW concept of   vorticity \cite{Ra87}, the vortical multipole operator
$\hat{M}(\rm{vor}; \:\lambda\mu;\: k)$ is obtained from (\ref{20}) by the substitution
\begin{equation}\label{24}
\left[\:\vec{\nabla} \times \hat{\vec{j}}_{\rm{nuc}}(\vec{r}) \:\right] \; \rightarrow \;
\hat{\vec{w}}(\vec{r}) = \vec{\nabla} \times \hat{\vec{j}}_{\rm{nuc}}(\vec{r}) -
\vec{\nabla} \times \hat{\vec{j}}_{\rm{irrot}}(\vec{r}).
\end{equation}
Using (\ref{19}), one gets in the long wavelength limit
\begin{eqnarray}\label{25}
&& \hat{M}(\rm{vor} ;\: \lambda\mu;\: k) = k \: \hat{M}(vor;\:\lambda \mu) + \ldots
\nonumber \\
 &&  = k \:\left[ \: \hat{M}(\rm{tor};\:\lambda \mu) + \hat{M}(com; \:\lambda \mu)\: \right] + \ldots
\end{eqnarray}
The toroidal term $\hat{M}(\rm{tor};\:\lambda \mu)$ is given by (\ref{23})
and the compressional operator is
\begin{eqnarray} \label{26}
\hat{M}(\rm{com} ;\:\lambda \mu)  = - k \: \hat{M'}(\rm{com}; \:\lambda \mu),
\end{eqnarray}
where $ \hat{M'}(\rm{com};\: \lambda \mu)$ is the familiar
compression multipole operator \cite{Sem81,Kv03}
\begin{equation} \label{27}
\hat{M'}(\rm{com};\:\lambda \mu) = \frac{1}{2\:(2 \lambda +3)}
\:\int \:d^3r \:\hat{\rho}(\vec{r}) \:
r^{\lambda+2} \: Y_{\lambda \mu}(\hat{r}).
\end{equation}
From the above equations it follows \cite{Kv11} that
the vorticity multipole operator has the form
\begin{eqnarray} \label{28}
&& \hat{M}(\rm{vor};\:\lambda \mu) = \hat{M}(\rm{tor};\:\lambda \mu)
+ \hat{M}(\rm{com}; \:\lambda \mu)
\nonumber \\
&& \quad =  - \frac{i}{c} \:\frac{1}{2\lambda+3}\:\sqrt{\frac{2\lambda+1}{\lambda+1}}\:
\int \: d^3r \:  \hat{\vec{j}}_{\rm{nuc}}(\vec{r})
\nonumber \\
&& \qquad \qquad \qquad \qquad \qquad \:\cdot \: r^{\lambda+1}
 \:\vec{Y}_{\lambda \: \lambda+1 \:\mu}(\hat{r}).
\end{eqnarray}
This definition shows that  the vortical
operator $\hat{M}(\rm{vor};\:\lambda \mu;\:k)$ (\ref{25})
is a first order correction to the leading   $E\lambda$ operator  (\ref{20}).

For the dipole transitions, the operators are to be accordingly
modified in order to remove the spurious
contribution induced by the center of mass motion. Following
\cite{Kv11}, only the toroidal (\ref{29}) and compressional (\ref{30})
operators are actually modified, while the vorticity operator (\ref{31})
remains unchanged.

\section{SRPA approach}

The SRPA Hamiltonian is self-consistently derived \cite{Ne02,Ne06} from the functional
\cite{Ben03}
\begin{equation}\label{32}
\mathcal{E} =
  \mathcal{E}_{\mathrm{kin}} + \mathcal{E}_{\mathrm{Sk}}+ \mathcal{E}_{\mathrm{pair}}
+ \mathcal{E}_{\mathrm{Coul}}
\end{equation}
involving kinetic-energy, Skyrme, pairing, and Coulomb terms. The
Skyrme functional $\mathcal{E}_{\mathrm{Sk}} (\rho, \tau, \vec{J},
\vec{j}, \vec{s}, \vec{T})$ \cite{Skyrme,Vau72,En75} depends on the time-even nucleon ($\rho$),
kinetic-energy ($\tau$), and spin-orbit ($\vec{J}$) densities as well as on the time-odd  current
($\vec{j}$), spin ($\vec{s}$), and vector kinetic-energy ($\vec{T}$) densities.

The Hamiltonian is composed of the HFB mean field and residual potential \cite{Ne02,Ne06}
\begin{equation}\label{33}
    \hat{H} = \hat{h}_{\mathrm{HFB}} + \hat{V}_{\mathrm{res}}.
\end{equation}
The HFB mean field is given by
\begin{equation}\label{34}
    \hat{h}_{\mathrm{HFB}} = \int d^3r \sum_{\alpha_+}
    [\frac{\delta \mathcal{E}}{\delta J_{\alpha_+}(\vec{r})}]
    \hat{J}_{\alpha_+},
\end{equation}
where $J_{\alpha_{+}}$ ($J_{\alpha_{-}}$) are time even (odd) densities and  $\hat
J_{\alpha_{\pm}}$ are the corresponding operators, with
$\alpha_+$ ($\alpha_-$) denoting the densities.

The  residual interaction is a sum of separable pieces
\begin{equation}\label{35}
    \hat{V}_{\mathrm res} = \frac{1}{2} \sum_{k, k' = 1}^{K}
    ( \kappa_{k k'}
    \hat{X}_k \hat{X}_{k'}+ \eta_{k k'} \hat{Y}_k \hat{Y}_{k'})
\end{equation}
with one-body fields
\begin{eqnarray}\label{36}
    \label{eq:operators}
    \hat{X}_k &=& i \int d^3r d^3r'
    \sum_{\alpha_+,\alpha'_+}
    \frac{\delta^2 \mathcal{E}}
    {\delta J_{\alpha_+}  \delta J_{\alpha'_+}}
    \langle[\hat{P}_k , \hat{J}_{\alpha_+} ]\rangle
    \hat{J}_{\alpha'_+} \; ,
\\ \nonumber
\hat{Y}_k &=& i \int \int d^3r d^3r'
    \sum_{\alpha_-,\alpha'_-}
    \frac{\delta^2 \mathcal{E}}
    {\delta J_{\alpha_-} \delta J_{\alpha'_-}}
    \langle[\hat{Q}_k , \hat{J}_{\alpha_-} ]\rangle
    \hat{J}_{\alpha'_-}
\end{eqnarray}
and strength constants
\begin{equation}\label{37}
  \kappa_{k k'}^{-1 } =
  - i \langle [\hat{P}_{k},{\hat X}_{k'}] \rangle \; , \quad
  \eta_{k k'}^{-1 }
  = -i
  \langle [\hat{Q}_{k},{\hat Y}_{k'}] \rangle \; .
\end{equation}
The fields $\hat{X}_k$ and $\hat{Y}_k$ are time-even and time-odd, respectively.
The corresponding generators are time-odd $\hat{P}_{k}=i[\hat H,\hat{Q}_{k}]$ and
time-even $\hat{Q}_{k}$.

The two-body potential includes all possible  separable terms allowed by the functional.
In practice, the  number of fields is determined by the number of  the generators
$\hat{Q}_k$ or $\hat{P}_k$ chosen as input.
Such a  choice   is crucial for generating a minimal set of one-body fields
 appropriate for the excitation mode to be studied and, therefore, render feasible
 the numerical implementation. An optimal set of   generators
were determined for $E1$ \cite{Ne02,Ne06} and  $M1$ \cite{Ve09,Ne10} modes.
The set of generators needed for   the modes under investigation here  was
discussed in the Appendix D of Ref. \cite{Kv11}. The Skyrme separable RPA may
be also derived by an alternative way \cite{NVG_98,Sever_08}.
\begin{figure}[t]
\includegraphics[width=7.1cm]{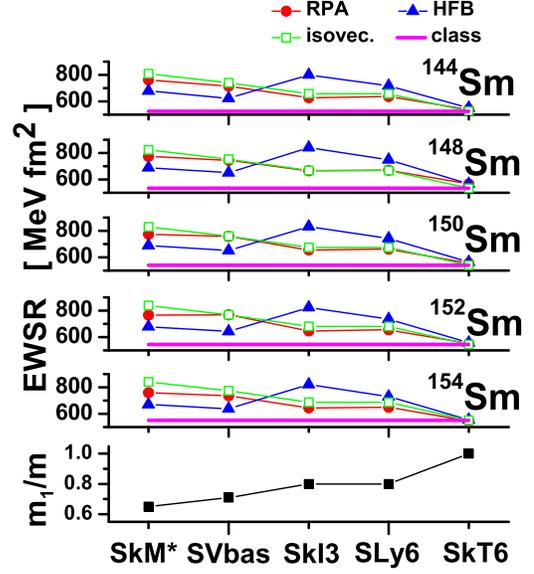}
\vspace{5mm}
\caption{ (color online) The HF/HFB (blue curve with triangles)
and RPA (red curve with circles) EWSR  in  $^{144-154}$Sm
isotopes,  calculated with different Skyrme forces. For the
comparison, the TRK values computed with the bare nucleon mass $m$
(pink horizontal line) and isovector effective mass $m_1$ (green
curve with rectangles) are depicted. The values of $m_1/m$ are
given in the low panel.}
\end{figure}

The self-consistent separable form (\ref{35}) of the residual interaction enables us
to transform the RPA equations of motion into a system of homogeneous algebraic
equations in the  forward $c_{ij}^{(\nu q -)}$ and backward $c_{ij}^{(\nu q +)}$
amplitudes of the RPA phonon creation operators \cite{Ne02}
\begin{equation} \label{38}
\hat{C}^{+}_{\nu} = \sum_{q=n,p} \;\sum_{ij\in q} \:
\left(c_{ij}^{(\nu q-)}\:\alpha^{+}_{i} \alpha^{+}_{j}
- c_{ij}^{(\nu q+)}\:\alpha_{j} \alpha_{i}\:\right),
\end{equation}
where $\alpha^{+}_{i}$ and $\alpha_{i}$ are quasiparticle creation and annihilation
operators, respectively. The rank of the matrix of the system of algebraic equations
is 4K,  where K is the number of separable terms in (\ref{35}). Usually
$K=3-5$ is sufficient to describe a given mode \cite{Ne02,Ve09}. So low rank
allows very efficient and fast solution of  the SRPA equations. They yield
one-phonon states $\mid \nu\rangle = \hat{C}^{+}_{\nu} \:|\rm{RPA}\rangle$
of energy $E_{\nu}$, with $|\rm{RPA}\rangle$ being the phonon vacuum.

\section{Thomas-Reiche-Kuhn sum rule}

 In Fig. 10, the integral
HFB and RPA energy-weighted strengths (\ref{43}) are compared to
the TRK sum rule for the Skyrme forces \cite{Ri80,Li89}
\begin{equation} \label{47}
\text{EWSR}_{\text{TRK}} = \frac{9\:\hbar^2}{8\pi
m_1}\:\frac{N\!Z}{A} \;.
\end{equation}
Unlike the common case \cite{Ri80}, this expression includes not
the bare nucleon mass $m$ but the isovector effective mass $m_1$
so as to take into account the contribution of the
velocity-dependent terms (VDT) of the Skyrme forces, see an
extensive discussion  in Refs.
\cite{Nest_IJMPE_08,Li89}. It is seen that,
for all Skyrme forces, the RPA strengths follow closely the TRK
values. The agreement takes place both for $m_1/m <$1 (SkM*,
SVbas, SkI3, SLy6 with active VDT) and $m_1/m =$1 (SkT6 with zero
VDT). This implies that i) $m_1$ successfully incorporates  almost
all VDT contributions to the EWSR, ii) our configuration space is
sufficiently large. For most of the forces, the HFB and RPA values
overestimate by 20-30\% the TRK values evaluated with the bare
nucleon mass $m$. This indicates an important role of the VDT.
Besides, the VDT causes a substantial difference between HFB and
RPA results.

Note that unlike the TRK sum rule (\ref{47}), the isoscalar EWSR
for the ISGDR \cite{Ha01} (\ref{Har_EWSR}) uses the bare mass $m$.
Following \cite{Li89}, the VDT contribute to both mean field and
residual interaction but, for T=0 RPA calculations, these
contributions almost compensate each other. Thus we may use the
bare mass $m$. Instead, for T=1 RPA calculations considered above,
the compensation does not take place, which results in using $m_1$
in (\ref{47}).

\end{document}